\documentclass[aps,twocolumn]{revtex4}
\usepackage{amsmath,amssymb}
\usepackage{graphicx}
\usepackage{epsfig}
\usepackage{graphicx}
\usepackage{epstopdf}

\newcommand{\be}{\begin{equation}}
\newcommand{\ee}{\end{equation}}
\newcommand{\bea}{\begin{eqnarray}}
\newcommand{\eea}{\end{eqnarray}}
\newcommand{\ba}{\begin{array}}
\newcommand{\ea}{\end{array}}

\newcommand{\eq}[1]{(\ref{#1})}

\begin{document}

\title{Information Causality and Noisy Computations}
\author{Li-Yi Hsu$^{a}$\footnote{lyhsu@cycu.edu.tw}, I-Ching Yu$^{b}$\footnote{896410029@ntnu.edu.tw}, Feng-Li Lin$^{b}$\footnote{linfengli@phy.ntnu.edu.tw, the corresponding author}}
\affiliation{${}^{a}$ Department of Physics, Chung Yuan Christian University, Chung-li
32023, Taiwan}
\affiliation{${}^{b}$ Department of Physics, National Taiwan Normal University,
Taipei, 116, Taiwan}

\begin{abstract}
We reformulate the information causality in a more general framework by
adopting the results of signal propagation and computation in a noisy
circuit. In our framework, the information causality leads to a broad class
of  Tsirelson inequalities. This fact allows us to subject information causality
to experimental scrutiny. A no-go theorem for reliable nonlocal
computation is also derived. Information causality prevents any physical
circuit from performing reliable computations.
\end{abstract}

\startpage{1}
\endpage{2}
\maketitle

\section{Introduction} 
As a physical theory, quantum mechanics has been
extremely successful in describing the microscopic physics. Nevertheless, its current framework is incapable of explaining the nature of quantum entanglement. Attempts to remedy this situation have been made by reconstructing quantum mechanics in terms of physical principles.
These physical principles should be able to yield or constrain the non-local correlation implied by the quantum entanglement. One such candidate is the principle of space-time causality.  This principle will constrain the possible non-local correlation such that any physical theory must be no-signaling \cite{pr},  i.e., signal cannot be send in the way of violating causality. However, a broad class of no-signaling theories other than quantum mechanics exist. Certain features, usually thought of as specifically quantum, are common for many of these theories \cite{s,gen}.  Clearly, no-signaling is insufficient as a principle to single out quantum mechanics.

Some of these theories are allowed to have  more non-local correlation than quantum mechanics \cite{non,gen,s1,s,sec}. Specifically, the  non-local
correlation in these theories can violate Bell-type inequalities by more than
Tsirelson's bound \cite{Tsirelson,pr}. From this perspective, we should
search for a physical principle as follows. The principle can single out
Tsirelson's bound as a limitation on the extent of the allowed correlation
for a physical theory. With the advent of quantum information science, some
principles of information theoretic flavor have been proposed. These proposed
candidates set the constraints on the physically realizable correlations. In
this Letter, we focus on a promising candidate --- the information causality.
Information causality states that, in a bipartite code protocol prepared
with any physically local or non-local resources, the accessible information
gain cannot exceed the amount of classical communication. In \cite{IC} information causality is demonstrated by a generic task similar to
random access codes (RAC) and oblivious transfer. In this task, a database
of $k$ bits is prepared:  $\vec{a}:=(a_{0},a_{1},\cdots ,a_{k-1})$, where each $a_{i}$ is a
random variable, which is only known by the first party, Alice. A second,
distant party, Bob, is given a random variable $b\in (0,\cdots ,k-1)$ along
with a bit $\alpha $ send by Alice. With the bit $\alpha $ and the
pre-shared correlation with Alice, Bob's task is to optimally guess the bit $a_{b}$. Then, according to information causality the quantity $I$ has an upper bound
\begin{equation}
I=\sum_{i=0}^{k-1}I(a_{i};\beta |b=i)\leq 1\;.  \label{ic-1}
\end{equation}
Here $I(a_{i};\beta |b=i)$ is the Shannon mutual information between $a_{i}$
and Bob's guessing bit $\beta $  under the condition $b=i$. Classically, $I$ can
reach 1 once $\alpha =a_{i}$ and $I(a_{i};a_{j})=\delta _{ij}$ ( i.e., the
Kronecker delta).

To perform the RAC task, Alice and Bob can use (earlier prepared and
distributed) correlations among either classical or quantum systems. These no-signaling correlation resources can be simulated by the no-signaling box (NS-box). The NS-box correlates the inputs and outputs of Alice and Bob in an imperfect way subjected to the probabilistic noise.
The noise of NS-box is intrinsically inherited from the underlying physical theory such as quantum mechanics. The quantity $I$ in \eq{ic-1} is unavoidably affected by the intrinsic noise of NS-box. In this framework, the signal decay theorem in \cite{signal,nc} for a noisy circuit is exploited to yield a tight bound for $I(a_b;\beta|b)$ in terms of noise of NS-box.   According to information causality,  the tight bound should also obey the upper bound in \eq{ic-1}.  By expressing the tight bound in terms of correlation functions between Alice's and Bob's measurement outcomes, this then yields our main result --- a broad class of  multi-setting Tsirelson-type inequalities.  As a result, we can then subject the physical principle of information causality to scrutiny by experimentally verifying or falsifying the generalized Tsirelson's bounds.

Without classical communication, the RAC can be regarded as  nonlocal computation. Therein, distant Alice and Bob compute a general Boolean function without knowing the other's input. Here, NS-box can be regarded as a noisy gate for non-local computation \cite{nl}.  Noise of the gate is closely related to the reliability of non-local computation. The computational noise of the gate is related to the intrinsic reliability of the physically realized NS-box. In this aspect, we can tackle a fundamental question on noisy computation with its nonlocal version. As raised by von Neumann \cite{vonN}, this question is originally stated as follows. Could physical circuits of finite size perform the reliable noisy non-local computation of any Boolean function? Based on constraint by the information causality for any physical circuit, we will see that the answer is negative in non-local computation.

  The paper is organized as follows. In the next section we derive the Tsirelson-type inequalities from the information causality by using the theorem of signal propagation. In the section III we discuss the implication of information causality on the nonlocal quantum computation and yield a no-go theorem for reliable nonlocal quantum computation. Finally we briefly conclude our paper in section IV.  Moreover, in the Appendix we give the details of verifying our newly-derived Tsirelson-type inequalities by using the method of semidefinite programing.

\section{Tsirelson-type inequalities }
We start by reformulating the NS-box
as a noisy distributed gate for nonlocal computation. The NS-box is
initially distributed between two distant parties, Alice and Bob. Locally, Alice and
Bob input bit strings $\vec{x}$ and $\vec{y}$, respectively, into half of the box, which
then outputs bits $A_{\vec{x}}$ and $B_{\vec{y}}$, respectively.  The lengths
of the bit strings can be chosen by design. Our NS-box is further characterized by the conditional joint
probabilities $\Pr \left[ A_{\vec{x}}+B_{\vec{y}}=f(\vec{x},\vec{y})\;|\vec{x%
},\vec{y}\right] $. Therein, $f(\vec{x},\vec{y})$ is the task function.
Notably, if the NS-box is physically realizable, these joint
probabilities must fulfill the no-signaling conditions.

In the RAC protocol, the chosen task function $f(\vec{x},\vec{y})$ depends on
how we encode Alice's database $\vec{a}$ and Bob's given random variable $b$
into $\vec{x}$ and $\vec{y}$, respectively. From now on, we will implicitly use the following protocol.  Firstly, Alice encodes her database $\vec{a}$ into the $(k-1)$-bit string $\vec{x}:=(x_1,\cdots, x_{k-1})$ by $x_{i}=a_{0}+a_{i}$. Alice's half of NS-box then produces an outcome  $A_{\vec{x}}$.  At the same time,  Bob encodes the given $b$ in to $(k-1)$-bit string $\vec{y}:=(y_1,\cdots,y_{k-1})$ by $y_i=\delta_{b,i}$ for $b\ne 0$, and $\vec{y}=\vec{0}$ for $b=0$.   Bob's half of NS-box then produces an outcome  $B_{\vec{y}}$.  Secondly, Alice sends Bob a bit $\alpha=a_0+ A_{\vec{x}}$. The optimal strategy for Bob's task is to  output  a guess bit $\beta=\alpha+B_{\vec{y}}$.  As a result, Bob can decode Alice's bit $a_{b}$ successfully whenever  $A_{\vec{x}}+B_{\vec{y}}=\vec{x} \cdot \vec{y}$ (modulo $2$) is true.  Most of the calculations in this Letter are modulo-2 defined.

In quantum mechanics, Alice's and Bob's outcomes can be produced by performing the corresponding measurement of $2^{k-1}$ and $k$ settings, respectively.  For the above protocol,  the success probability of Bob's task in guessing Alice's bit $a_{b}$ is related
to the one for noisy computation as follows
\begin{equation}
\Pr[\beta=a_{b}|  b\;]=\frac{1}{N_{\vec{x}}}\sum_{\{\vec{x}\}}
\Pr\left[  A_{\vec{x}}+B_{\vec{y}}=f(\vec{x},\vec{y})|\vec{x},\vec
{y}\;\right] ,
\end{equation}
where $N_{\vec{x}}$ is the cardinality of the input space spanned
by the encoding $\{\vec{x}\}$. By defining the correlation functions between Alice's and Bob's measurement outcomes as $C_{\vec{x},\vec{y}
}:=\sum_{A_{\vec{x}}=0,1}\sum_{B_{\vec{y}}=0,1}(-1)^{A_{\vec{x}}+B_{\vec{y}}}\Pr\left[
A_{\vec{x}},B_{\vec{y}}\;|\vec{x},\vec{y}\;\right]  $,  we find
\begin{equation}
\xi_{\vec{y}}=\frac{1}{N_{\vec{x}}}\sum_{\{\vec{x}\}}(-1)^{f(\vec
{x},\vec{y})}C_{\vec{x},\vec{y}}. \label{xitoC}
\end{equation}
where the coding noise parameter is defined as $\xi_{\vec{y}}:=2\Pr\left[  \beta=a_{b}\;|
b\;\right]  -1$. The sub-index $\vec{y}$ of $\xi_{\vec{y}}$ is understood to be equivalent to Bob's given parameter $b$ via encoding.

One of the main results of this paper is a broad class of Tsirelson's bound implied by information causality, i.e.,
\begin{equation}
|\sum_{\{\vec{y}\}}\xi_{\vec{y}}\;|=\frac{1}{N_{\vec{x}}}|\sum
_{\{\vec{x}\},\{\vec{y}\}}(-1)^{f(\vec{x},\vec{y})}C_{\vec{x},\vec{y}}
\;|\leq\sqrt{k}\;. \label{tsirelson}
\end{equation}
For $k=2$, it is easy to check that \eq{tsirelson} is the Tsirelson's bound $|C_{0,0}+C_{0,1}+C_{1,0}-C_{1,1}|\leq2\sqrt{2}$ \cite{Tsirelson}.  For the case of $k>2$ with $f(\vec{x},\vec{y})=\vec{x} \cdot \vec{y}$, we have verified \eq{tsirelson} to be the Tsirelson's bound in quantum mechanics by using the semidefinite programing \cite{SDP}.  Please see Appendix for more detailed discussions.

Indeed, later we will see that information causality will render (\ref{tsirelson}). This implies that  information causality can be tested by experimental verification or refutation via the measurement of the correlation functions of a quantum system.

  In order to arrive the Tsirelson's bound \eq{tsirelson} from the information causality constraint \eq{ic-1}, we need to relate  $I(a_b,\beta|b)$ to $\xi_{\vec{y}}$. It turns out that this can be done by using the following signal decay theorem on the signal propagation \cite{signal,nc}.

\textit{Theorem 1:} Let $X$, $Y$ and $Z$ be Boolean random variables.
Consider a cascade of two communication channels: $X\hookrightarrow
Y\hookrightarrow Z$. $X$ and $Y$ are the input and the output of the
first channel, respectively. Let $Y$ in turn be the input of a cascading
binary symmetric channel $C_{\epsilon}$ with a noise parameter $\epsilon$,
i.e.,
\begin{equation}
C_{\epsilon}=\left(
\begin{array}{cc}
\frac{1}{2}(1+\epsilon) & \frac{1}{2}(1-\epsilon) \\
\frac{1}{2}(1-\epsilon) & \frac{1}{2}(1+\epsilon)
\end{array}
\right) .  \notag  \label{ek}
\end{equation}
Let $Z$ be the output of $C_{\epsilon}$, (i.e., $Z=\overline{Y}$ with the
bit-flipping probability $\frac{1}{2}(1-\epsilon)$)
\begin{equation}
{\frac{I(X;Z)}{I(X;Y)}}\leq\epsilon^{2}.
\end{equation}
A special case arises if the first channel is noiseless or trivial, i.e., $
I(X;Y=X)=1$ such that $I(X;Z)\leq\epsilon^{2}$. Note also that regardless of the properties of the second channel, there is a data processing inequality $I(X;Z)\leq I(X;Y)$.

We apply this theorem to our RAC protocol as follows. Because Alice's  database $a_{0},a_{1}, \cdots, a_{k-1}$ are random variables and independent of
each other, so that all the $a_{j}$'s with $j\neq i$ can be fixed without
disturbing $I(a_{i}\;;\beta |b)$. Let $X=a_{i}$, $Y=a_{0}+f(\vec{x},\vec{y})$
, and $Z=\beta $. Here $Y$ is Bob's ideal answer and hence $I(X;Y)=1$. The
coding noise $\epsilon $ for our protocol is $\xi _{\vec{y}}$, then
according to the Theorem 1, we have
\begin{equation}
I(a_{i}\;;\beta |b=i)\leq \xi _{\vec{y}}^{2}.  \label{ub}
\end{equation}

Therefore, the information causality in Eq. (\ref{ic-1}) yields
\begin{equation}
I\leq \sum_{\{\vec{y}\}}\xi _{\vec{y}}^{2}\leq 1.  \label{quadratic}
\end{equation}
In \cite{IC}, similar inequalities are derived to avoid the divergence
of $I$, which justifies the information causality. However, such trouble
does not exist in our reformulation because of the tight bound of Theorem 1. With
the help of (\ref{xitoC}) the second inequality in (\ref{quadratic}) becomes
a quadratic Tsirelson-type inequality for the correlation function $C_{\vec{x},\vec{y}}$. Moreover, using the Cauchy-Schwarz inequality, we can obtain $|\sum_{\{\vec{y}\}}\xi _{\vec{y}}|\leq \sqrt{k}$, which results in the linear Tsirelson inequality of Eq. (\ref{tsirelson}).

\section{Noisy nonlocal computation}
In the previous discussion we have considered
the information causality using a single nonlocal NS-box. Instead, we can
treat the NS-box as a  non-local gate for performing the nonlocal computation, i.e.,
computing the function $f(\vec{x},\vec{y})$ \cite{nl}. Unlike using the same
gate for the RAC, no classical communication between Alice and Bob is required
to perform the nonlocal computation.
In details, Alice's and Bob's local  outputs are $A_{\vec{x}}$
and $B_{\vec{y}}$, respectively. The computation is successful if $A_{\vec{x}}+B_{\vec{y}}=f(\vec{x},\vec{y})$. The computational noise parameter is
defined as
\be
\epsilon _{\vec{x},\vec{y}}:=2\Pr [A_{\vec{x}}+B_{\vec{y}}=f(\vec{x},\vec{y})|\vec{x},\vec{y}]-1.  \label{spi}
\ee
From (\ref{spi}) and (\ref{xitoC}) the computational noise of the gate is
related to its coding noise by
\be\label{codetocomp}
\xi_{\vec{y}}=\frac{1}{N_{\vec{x}}}\sum_{\{\vec{x}\}}\epsilon_{\vec{x},\vec{y}}\;.
\ee
Basically, computational errors inherently come from the gate noise.
Information causality constraints the noisy extent of the NS-box as a gate.
From this perspective, information causality is deeply connected with
nonlocal computation.
\begin{figure}[th]
\includegraphics[width=0.9\columnwidth]{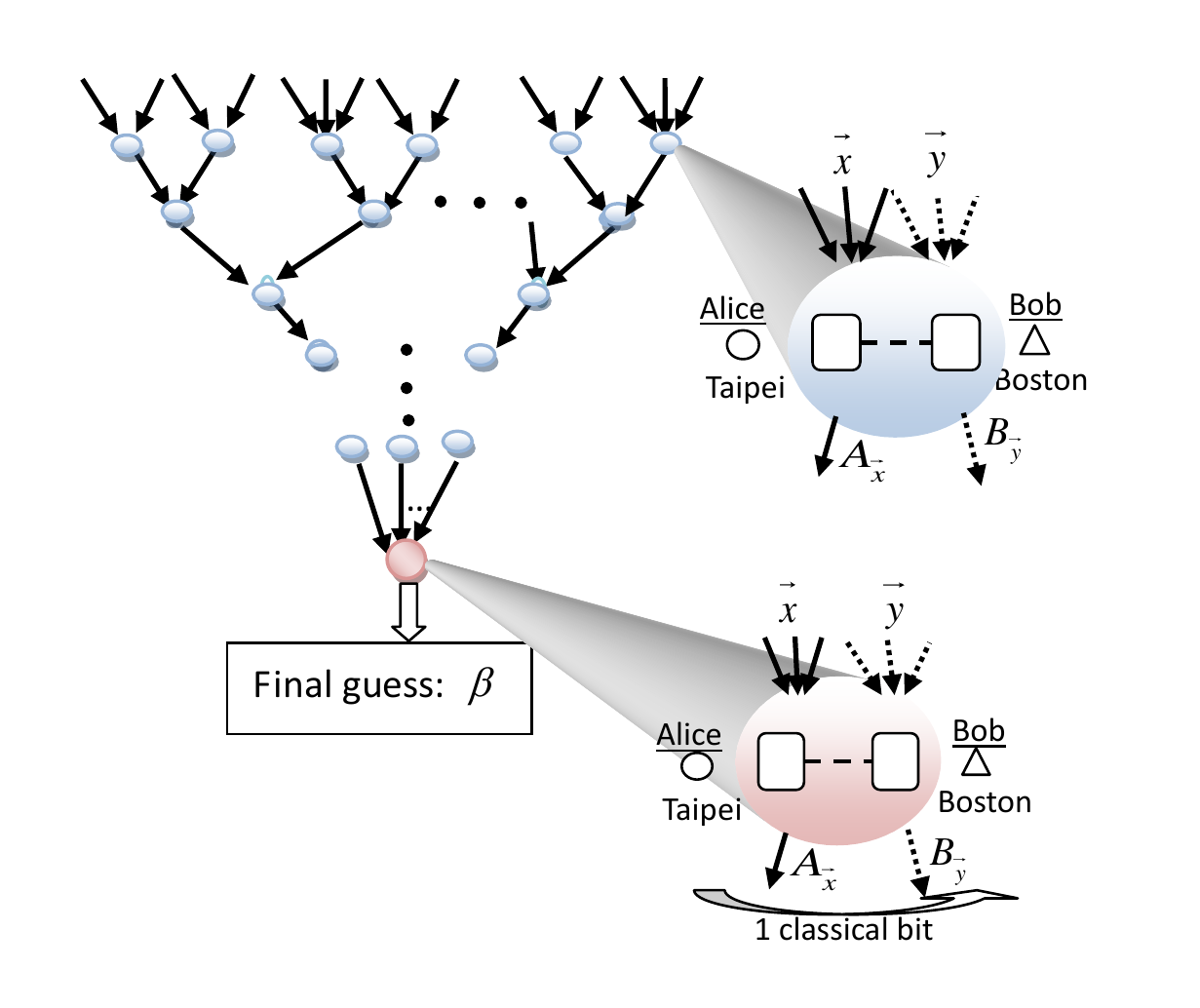}\newline
\caption{RAC protocol for a $(n,k,l)$-circuit. Each vertex of the circuit
corresponds to a NS-box, with its details shown in the big ellipses.}
\label{tree}
\end{figure}

Furthermore, we can combine the NS-box gates to form a more complicated circuit without worrying about the coding protocol. Then the total task function for the whole circuit will be a
complicated function, i.e., a composite of task functions of all NS-boxes. We
can then try to answer the following fundamental question: \textit{could a
noiseless (nonlocal) computation be simulated using a noisy nonlocal
physical resource?}

 Specifically, we consider the so-called $(n$, $k$, $l)$-circuit, $G$, formed by
cascading layers of noisy gates into a circuit in the form of a directed,
acyclic tree (see Fig 1). On the top of $G$, there are $n$ inputs to the
NS-boxes --- the leaves; at the bottom there is only one NS-box --- the root. The
longest path from the leaves to the root is called the depth of the circuit,
denoted by $l$. The maximum input number of a gate in $G$ is $k$. Note
that, in \cite{IC} $G$ comprises $k=2$ gates and is exploited to compress $n$
bits of $\vec{x}$ into one bit $A_{\vec{x}}$. However, there is no restriction
on the task function for each NS-box, as long as the final circuit is a
consistent acyclic tree diagram.

We then use the circuit $G$ to perform the following nonlocal computation.
Alice's $n$-bit database $\vec {a}:=(a_{0},a_{1},\cdots,a_{n-1})$ is given
to the leaves of $G$, and a conditional input $b\in\{0,1,\cdots ,n-1\}$ is
given to the distant Bob. The previous encoding $\vec{a}\rightarrow \overrightarrow{x}$ and $b\rightarrow$ $\vec{y}$ for the RAC protocol, is also exploited here. Alice's output is properly encoded and then fed into the NS-box at the next layer, again with Bob's conditional input. The
same procedure is performed recursively until reaching the root, with its output as the
answer to the total task function at the root.

Alternatively, Bob's decoding gates can be thought to be noise free, and the computational noise
is only due to Alice's encoding gates, and vice versa. This makes it easier to
understand the above procedure of noisy computation.  Now we can consider the information flow of $G$.

\textit{Theorem 2: } For a noisy, local circuit $G$ with an arbitrary
depth, the root outputs at most one-bit information.

Note that the circuit $G$ can perform the RAC if the appropriate protocol is
given at each layer and 1-bit communication is allowed for the whole
process. Then, the above theorem implies that information causality
holds true for the circuit $G$.

To prove the theorem, we will show that the mutual information between
the leaves and the root of $G$ is bounded by one. This can be done by mathematical
induction as follows. We begin with a circuit of depth one, which is nothing
but a single NS-box; information causality ensures the bound. We then
assume that the bound holds true for a circuit of depth $\ell$. According to
information causality and sub-additivity, the mutual information $I_{\ell}^{(m)}$ between the leaves and the root obeys $I_{\ell}^{(m)}\leq \sum_{i_{m}}I(X_{i_{m}};R_{m})=\sum_{i_{m}}I(X_{i_{m}};R_{m}|$Bob's knowledge$)\leq1$, where the index $m$ labels a collection of circuits of depth $\ell$ with root $R_{m}$, and the index $i_{m}$ labels the inputs of the $m$-th circuit. Now, we construct a circuit of depth $\ell+1$ by connecting all roots $R_{m}$'s to a single NS-box whose output is $R$. Then, the mutual information $I_{\ell+1}$ between leaves and root $R$ of the final circuit should obey the subadditivity, i.e., $I_{\ell+1}\leq\sum_{m} \sum_{i_{m}}I(X_{i_{m}};R)$. From Theorem 1, we have $I(X_{i_{m}};R)\leq \xi_{m}^{2}I(X_{i_{m}};R_{m})$ because we have a cascade of two channels: $X_{i_{m}}\hookrightarrow R_{m}\hookrightarrow R$ where the second channel is a binary symmetric one with the noise $\xi_{m}$. Using this result, we have $I_{\ell+1}\leq\sum_{m}\xi_{m}^{2}\sum_{i_{m}}I(X_{i_{m}};R_{m})\leq\sum_{m} \xi_{m}^{2}\leq1$. Q.E.D.

Here, we have only considered the case in which the computational noise is isotropic to $\vec{x}$, denoted by $\epsilon _{\vec{y}}$. From (\ref{codetocomp}) we have $\epsilon _{\vec{y}}=\xi _{\vec{y}}$ and the information causality requires $\sum_{\{\vec{y}\}}\epsilon _{\vec{y}}^{2}\leq 1$. We would like to know whether the reliable computation is also constrained by the information
causality or not. To check this, we invoke the main Evans-Schulman theorem
on the conditions for  reliable noisy computation as follows \cite{signal,nc}.

{\it Evans-Schulman Theorem:} A circuit of complete $k$-ary tree with depth $l$ ( i.e., $n=k^{l}$) can perform $\delta $-reliable noisy computation only
\begin{itemize}
\item (i) if $\sum_{\{ \vec{y} \}} \epsilon^{2}_{\vec{y}}>1$ then $\ell\ge
\log(n\Delta)/\log(\sum_{\{ \vec{y} \}} \epsilon^{2}_{\vec{y}})$\;,

\item (ii) if $\sum_{\{\vec{y}\}}\epsilon_{\vec{y}}^{2}\leq1$ then $n\leq1/\Delta$,
\end{itemize}
where $\Delta:=1+\delta\log\delta+(1-\delta)\log(1-\delta)$. The computation
is called $\delta$-reliable if the root outputs correctly with a probability $1-\delta$ (with $\delta<1/2$). This theorem provides stricter conditions than the original proposal by Von Neumann \cite{vonN,Pippenger}.

By definition, smaller $\epsilon_{\vec{y}}$ means larger noise, and the
condition (ii) is for the cases with larger noise such that only functions
with a smaller number of inputs can be reliably computed. Immediately, we see
that information causality implies a large computational noise for the RAC
circuit such that only condition (ii) for reliable noisy computation can
possibly be fulfilled. As a result, Alice's output asymptotically becomes random because $\Delta\rightarrow 0$ and hence $\delta\rightarrow\frac{1}{2}$ as $n\rightarrow\infty$. In summary, this implies that \textit{information causality prevents any physically realizable $(n,k,l)$-circuit from achieving reliable computations of excessively complicated
functions, i.e., with either too many inputs or lengthy steps needed.}

 The above result applies only when classical communication between Alice and Bob is disallowed. Under such circumstances, the noise of the gate is intrinsically constrained by the underlying physical theory. Otherwise, the classical communication can be exploited to improve the reliability of the gates so that the no-go result could be lifted.

\section{Conclusion}  We show how information causality leads to Tsirelson bounds
in a much easier way.  A series of new Tsirelson bounds are then derived. These bounds provide
some playground to test the information causality by experiments, as done before to test the Bell inequality. 
Moreover, deep ramifications concerning non-local quantum computation are also found and discussed. Especially, the no-go theorem for the reliable nonlocal quantum computation deserves more study to clarify its physical implication. 

The authors acknowledge financial support from the NSC of Taiwan under Contract No. NSC.99-2112-M-033-007-MY3 and 97-2112-M-003-003-MY3. This work is partially supported by NCTS.

\section*{Appendix}

In this appendix, we write down the detail of getting the Tsirelson-type inequalities derived from IC, and also check these inequalities directly by semidefinite programing (SDP).

We review the the RAC protocol as follows. Alice has a database of $k$ bits
$a_{0},a_{1},,,a_{k-1}$ where $a_{i}$ $\in\{0,1\}$ is the random variable
$\forall i\in(0,\cdots,k-1)$. The distant Bob is given a random variable
$b\in(0,...k-1)$ and a bit $\alpha$ sent by Alice. Bob's task is to guess
$a_{b}$. Here we will consider the RAC protocol with different settings. Case
(a) is proposed in the main text. In case (b), Alice's and Bob's settings are
modified. In the following, Alice's input is denoted by an $N$-bit string
$\vec{x}=x_{1}\ldots x_{N}$. Let $x=1+{\displaystyle\sum_{i=1}^{N}}
2^{i-1}x_{i}$, $1\leq x\leq2^{N}$. Bob's input is denoted by $N$- bit string
$\vec{y}=y_{1}\ldots y_{N}$.\\

\begin{itemize}
\item {\textbf{Case (a)}}\\
 Here $N=k-1$, and $x_{i}=a_{0}+a_{i}$ $\forall
i\in\{1,...,k-1\}$. $y_{i}=\delta_{i,b}$ $\forall i\in\{1,...,k-1\}$, if
$b\neq0$. $\vec{y}=\vec{0}$ if $b=0$. Let $y=1+ {\displaystyle\sum_{i=1}^{N}} iy_{i}$, $1\leq y\leq k$. In this case, the Tsirelson-type inequality derived
from information causality following the procedure in the main text is
\begin{equation}
|\sum_{\{\vec{x}\},\{\vec{y}\}}(-1)^{\vec{x}\cdot\vec{y}}C_{\vec{x},\vec{y}
}|\leq2^{k-1}\sqrt{k}.\label{Casea}
\end{equation}\\

\item {\textbf{Case (b)} }\\
Here $N=k$, $x_{i}=a_{i-1}$, and $y_{i}=\delta_{i,b+1}$ $\forall i\in\{1,...,k\}$. Let 
$y={\displaystyle\sum_{i=1}^{N}}iy_{i}$, $1\leq y\leq k$. Then, the Tsirelson-type inequality from information
causality is
\begin{equation}
|\sum_{\{\vec{x}\},\{\vec{y}\}}(-1)^{\vec{x}\cdot\vec{y}}C_{\vec{x},\vec{y}
}|\leq2^{k}\sqrt{k}.\label{Caseb}
\end{equation}
\end{itemize}

We now briefly introduce the semidefinite programming \cite{SDP}. SDP is the problem of
optimizing a linear function subjected to certain conditions associated with a
positive semidefinite matrix $X$, i.e., $v^{\dag}Xv\geq0$, for
$v \in\mathbb{C}^{n}$, and is denoted by $X\succeq0$. It can be formulated as
the standard primal problem as follows. Given the $n\times n$ symmetric
matrices $C$ and $D_{q}$'s with $q=1,\cdots,m$, we like to optimize the
$n\times n$ positive semidefinite matrix $X\succeq0$ such that we can achieve the following:
\begin{subequations}
\label{cons}%
\begin{align}
minimize \qquad &  Trace(C^{T}X)\\
subject \quad to \qquad &  Trace(D_{q}^{T}X)=b_{q}, \quad q=1,\cdots,m\;.
\end{align}
\end{subequations}
Corresponding to the above primal problem, we can obtain a dual problem via a
Lagrange approach \cite{CV}. The Lagrange duality can be understood as the
following. If the primal problem is

\begin{subequations}
\begin{align}
minimize\qquad &  f_{0}(x)\\
s.t.\qquad     &  f_{q}(x)\leq0,\quad q\in1...m.\\
               &  h_{q}(x)=0,\quad q\in1...p,
\end{align}
\end{subequations}
the Lagrange function can be defined as%
\begin{equation}
L(x,\lambda,\nu)=f_{0}(x)+\Sigma_{q=1}^{m}\lambda_{q}f_{q}(x)+\Sigma_{q=1}%
^{p}\nu_{q}h_{q}(x), \label{Lf}%
\end{equation}
where $\lambda_{1}$,\ldots, $\lambda_{m}$, and $\nu_{1}$,\ldots,$\nu_{p}$ are
Lagrange multipliers respectively.
Due to the problem and (\ref{Lf}), the minima of $f_{0}$ is bounded by
(\ref{Lf}) under the constraints when $\lambda_{1}$,\ldots, $\lambda_{m}\geq
0$.
\[
\mathop{\inf}_{x}f_{0}\geq\mathop{\inf}_{x}L(x,\lambda,\nu).
\]
Then the Lagrange dual function is obtained.
\[
g(\lambda,\nu)=\mathop{\inf}_{x}L(x,\lambda,\nu).
\]
$g(\lambda,\nu)\leq p$ ($p$ is the optimal solution of $f_{0}(x)$ ), for
$\lambda_{1}$,\ldots, $\lambda_{m}\geq0$ and arbitrary $\nu_{1}$,\ldots
,$\nu_{p}$. The dual problem is defined.
\begin{subequations}
\label{Ldp}%
\begin{align}
maximize\qquad &  g(\lambda,\nu)\\
s.t.\qquad &  \lambda_{q}\geq0.\quad(q\in\{1...m\})
\end{align}
\end{subequations}
We can use the same method to define the dual problem for SDP. From the primal
problem of SDP (\ref{cons}), we can write down the dual function by using
minimax inequality \cite{DPP}.%
\begin{widetext}
\begin{align}
\mathop{\inf}_{X\succeq 0}Trace(C^{T}X)  &  =\mathop{\inf}_{X\succeq 0}Trace(C^{T}X)+\sum_{q=1}^{m}y_{q}(b_{q}-Trace(D_{q}^{T}%
X))\nonumber\\
&  =\mathop{\inf}_{X\succeq0}\mathop{\sup}_{y}\sum_{q=1}^{m}y_{q}%
(b_{q})+Trace((C^{T}-\sum_{q=1}^{m}y_{q}D_{q}^{T})X)\nonumber\\
&  \geq\mathop{\sup}_{y}\mathop{\inf}_{X\succeq 0}\sum_{q=1}^{m}y_{q}%
(b_{q})+Trace((C^{T}-\sum_{q=1}^{m}y_{q}D_{q}^{T})X)\nonumber\\
&  =\mathop{\sup}_{y}\mathop{\inf}_{X\succeq 0}\sum_{q=1}^{m}y_{q}%
(b_{q})+Trace((C-\sum_{q=1}^{m}y_{q}D_{q})^{T}X).
\end{align}
The optimal solution of dual function is bounded under some vector $y$ \cite{SDP2}.
\begin{align}
\mathop{\sup}_{y} \mathop{\inf}_{X\succeq 0}\sum_{q=1}^{m}
y_{q}(b_{q}) + Trace((C-\sum_{q=1}^{m}y_{q}D_{q})^{T}X)=  &  \mathop{\sup}_{y}%
\sum_{q=1}^{m} y_{q}(b_{q}) \qquad when \quad C-\sum_{q=1}^{m}y_{q}D_{q}
\succeq 0 \quad \nonumber\\
&  -\infty \qquad otherwise.
\end{align}
\end{widetext}
The corresponding dual problem is
\begin{subequations}
\begin{align}
maximize \qquad &  \sum_{q=1}^{m} y_{q}(b_{q})\\
s.t. \qquad &  S=C-\sum_{q=1}^{m}y_{q}D_{q} \succeq0.
\end{align}
\end{subequations}

If the feasible solutions for the primal problem and the dual problem attain
their minimal and maximal values denoted as $p^{\prime}$ and $d^{\prime}$
respectively, then $p^{\prime}\geq d^{\prime}$, which is called the duality
gap. This implies that the optimal solution of primal problem is bounded by
dual problem. This then leads to the following: Both the primal and the dual
problems attain their optimal solutions when the duality gap vanishes, i.e.,
$d^{\prime}=p^{\prime}$.

We now use SDP to check the Tsirelson-type bound.
To cast the above problem of
finding the Tsirelson's bound in the context of quantum mechanics, we need to
use Tsirelson's theorem \cite{T3}. It says that for any quantum state
$|\Psi\rangle\in\mathbb{A}\bigotimes\mathbb{B}$ shared by two observers Alice
and Bob with their measurement outcomes being $A_{x}\in\lbrack-1,1]$ and
$B_{y}\in\lbrack-1,1]$, respectively. The correlation function can be
expressed by the inner product of two real unit vectors $\alpha_{x}$,
$\beta_{y}\in\mathbb{R}^{t+v}$.\textbf{ }Therein, $t$ and $v$ are the numbers
of Alice's and Bob's measurement settings, respectively. In detail,
$C_{\vec{x},\vec{y}}$ used in (\ref{Casea}) or (\ref{Caseb}), the Tsirelson's
theorem guarantees that we have $C_{\vec{x},\vec{y}}=\alpha_{x}\cdot\beta_{y}%
$. Then, we can cast the problem of finding the Tsirelson bound in
(\ref{Casea}) or (\ref{Caseb}) into the following form of optimal problem for
SDP,
\begin{subequations}
\label{SDPT}%
\begin{align}
maximize\qquad &  |\sum_{\{\vec{x}\},\{\vec{y}\}}(-1)^{\vec{x}\cdot\vec{y}%
}\alpha_{x}\cdot\beta_{y}\mid\\
s.t.\qquad &  \Vert\alpha_{x}\parallel=\Vert\beta_{y}\parallel=1\;,\qquad
\forall\;x,y. \label{SDPT1}%
\end{align}
Then, the associated dual problem is
\end{subequations}
\begin{subequations}
\label{SDPTD}%
\begin{align}
minimize\qquad &  \sum_{q=1}^{m}y_{q}\\
s.t.\qquad &  S=\sum_{q=1}^{m}y_{q}D_{q}-C\succeq0.
\end{align}
\end{subequations}
We now will turn the problem (\ref{SDPT}) into the primal problem
(\ref{cons}) by constructing the matrices $X$, $C$ and $A_{i}$'s from the
unit vectors $\alpha_{x}$ and $\beta_{y}$. Following the way in \cite{Wehner},
the mapping is as follows. Define the matrix $P$ whose columns are vectors
$(\alpha_{1},...,\alpha_{t},\beta_{1},...\beta_{v})$. Then the SDSP matrix $X$
is given by $P^{T}P$, which can be put into the following block form

\[
X=\left(
\begin{array}
[c]{cc}%
E & F\\
G & H
\end{array}
\right)
\]
where the matrix elements of each block are $E_{ij}=\alpha_{i}\cdot\alpha_{j}%
$, $F_{ib}=\alpha_{i}\cdot\beta_{b}$ , $G_{aj}=\beta_{a}\cdot\alpha_{j}$ and
$H_{ab}=\beta_{a}\cdot\beta_{b}$ with $i,j=1,\cdots,t$ $(t=2^{N})$ and
$a,b=1,\cdots,v$ $(v=k)$. Note that $F$ and $G$ are used in (\ref{SDPT}), and
instead $E$ and $H$ are used in (\ref{SDPT1}). Therefore, we can write down
the matrices $C$ and $D_{q}$'s accordingly so that the problem (\ref{SDPT}) is
equivalent to the problem (\ref{cons}). It is easy to see that $C$ is a
matrix with only non-vanishing off-diagonal block of matrix elements given by
$(-1)^{\vec{x}\cdot\vec{y}}$, and $D_{q}$'s are the diagonal matrices with
$(D_{q})_{st}=\delta_{s,q}\delta_{t,q}$. We omit their detailed forms here.

We take $k=2$ and $k=3$ in case(a) for example.

\begin{itemize}
\item \textbf{k=2}

Here $\vec{x}=x_{1}$\ and $\vec{y}=y_{1}$. According Eq. (\ref{Casea}), we
want to maximize $|C_{0,0}+C_{0,1}+C_{1,0}-C_{1,1}|$. Using the Tsirelson
theorem, it is equivlaent to maximizing $\alpha_{1}\cdot\beta_{1}+\alpha
_{1}\cdot\beta_{2}+\alpha_{2}\cdot\beta_{1}-\alpha_{2}\cdot\beta_{2}$. Such
Tsirelson bound has been showed by Wehner \cite{Wehner} using SDP. We just
show the numerical result. For more details, please see \cite{Wehner}. After
using SeDuMi program \cite{SDM} to solve SDP, the optimal for both primal and
dual problem is $2.8284$. It is consistent with the Tsirelson bound \cite{Tsirelson}
($2\sqrt{2}$) for the case two settings per site.

\item \textbf{k=3}
Here $\vec{x}=x_{1}x_{2}$ and $\vec{y}=y_{1}y_{2}$. Notably, $\vec{y}%
\in\{00,10,01\}$. The problem which we want to solve is
\begin{widetext}
\begin{align}
&  & maximize \quad|C_{00,00}+C_{00,10}+C_{00,01}+C_{01,00}+C_{01,10}%
-C_{01,01}\nonumber\label{ICT3a}\\
&  & +C_{10,00}-C_{10,10}+C_{10,01}+C_{11,00}-C_{11,10}-C_{11,01}|\nonumber\\
&  & =maximize \quad\alpha_{1}\cdot\beta_{1}+\alpha_{1}\cdot\beta_{2}%
+\alpha_{1}\cdot\beta_{3}+\alpha_{3}\cdot\beta_{1}+\alpha_{3}\cdot\beta
_{2}-\alpha_{3}\cdot\beta_{3}\nonumber\\
&  & +\alpha_{2}\cdot\beta_{1}-\alpha_{2}\cdot\beta_{2}+\alpha_{2}\cdot
\beta_{3}+\alpha_{4}\cdot\beta_{1}-\alpha_{4}\cdot\beta_{2}-\alpha_{4}%
\cdot\beta_{3}.
\end{align}
\end{widetext}
The $X$ matrix for primal problem is $X=S^{T}S$ where the columns of $S$
correspond the unit vectors $(\alpha_{1},\alpha_{2},\alpha_{3},\alpha
_{4},\beta_{1},\beta_{2},\beta_{3})$.%
\begin{widetext}
\[
X= \left(  {%
\begin{array}
[c]{ccccccc}%
\alpha_{1}\cdot\alpha_{1} & \alpha_{1}\cdot\alpha_{2} & \alpha_{1}\cdot
\alpha_{3} & \alpha_{1}\cdot\alpha_{4} & \alpha_{1}\cdot\beta_{1} & \alpha
_{1}\cdot\beta_{2} & \alpha_{1}\cdot\beta_{3}\\
\alpha_{2}\cdot\alpha_{1} & \alpha_{2}\cdot\alpha_{2} & \alpha_{2}\cdot
\alpha_{3} & \alpha_{2}\cdot\alpha_{4} & \alpha_{2}\cdot\beta_{1} & \alpha
_{2}\cdot\beta_{2} & \alpha_{2}\cdot\beta_{3}\\
\alpha_{3}\cdot\alpha_{1} & \alpha_{3}\cdot\alpha_{2} & \alpha_{3}\cdot
\alpha_{3} & \alpha_{3}\cdot\alpha_{4} & \alpha_{3}\cdot\beta_{1} & \alpha
_{3}\cdot\beta_{2} & \alpha_{3}\cdot\beta_{3}\\
\alpha_{4}\cdot\alpha_{1} & \alpha_{4}\cdot\alpha_{2} & \alpha_{4}\cdot
\alpha_{3} & \alpha_{4}\cdot\alpha_{4} & \alpha_{4}\cdot\beta_{1} & \alpha
_{4}\cdot\beta_{2} & \alpha_{4}\cdot\beta_{3}\\
\beta_{1}\cdot\alpha_{1} & \beta_{1}\cdot\alpha_{2} & \beta_{1}\cdot\alpha_{3}
& \beta_{1}\cdot\alpha_{4} & \beta_{1}\cdot\beta_{1} & \beta_{1}\cdot\beta_{2}
& \beta_{1}\cdot\beta_{3}\\
\beta_{2}\cdot\alpha_{1} & \beta_{2}\cdot\alpha_{2} & \beta_{2}\cdot\alpha_{3}
& \beta_{2}\cdot\alpha_{4} & \beta_{2}\cdot\beta_{1} & \beta_{2}\cdot\beta_{2}
& \beta_{2}\cdot\beta_{3}\\
\beta_{3}\cdot\alpha_{1} & \beta_{3}\cdot\alpha_{2} & \beta_{3}\cdot\alpha_{3}
& \beta_{3}\cdot\alpha_{4} & \beta_{3}\cdot\beta_{1} & \beta_{3}\cdot\beta_{2}
& \beta_{3}\cdot\beta_{3}%
\end{array}
} \right)
\]
\end{widetext}

According to (\ref{ICT3a}), the matrix $C$ is defined
\[
C=\frac{-1}{2}\times\left(  {%
\begin{array}
[c]{ccccccc}%
0 & 0 & 0 & 0 & 1 & 1 & 1\\
0 & 0 & 0 & 0 & 1 & -1 & 1\\
0 & 0 & 0 & 0 & 1 & 1 & -1\\
0 & 0 & 0 & 0 & 1 & -1 & -1\\
1 & 1 & 1 & 1 & 0 & 0 & 0\\
1 & -1 & 1 & -1 & 0 & 0 & 0\\
1 & 1 & -1 & -1 & 0 & 0 & 0
\end{array}
} \right)  .
\]
The norm of the vectors $(\alpha_{1},\alpha_{2},{\alpha_{3},\alpha_{4}%
,\beta_{1},\beta_{2}},\beta_{3})$ must be one is the source of the constrain.
Each of the matrix $D_{q}$ ($q=1...7$) is a $7\times7$ diagonal matrix with
the $q$-th diagonal element being one and zero others. The value $b_{q}$
($q=1...7$) is one.
The numerical result shows that the tight bound is $6.9282$, which essentially
agrees with (\ref{Casea}). When we get the optimal solution, the correlation
function matrix is
\begin{widetext}
\[
X=\left(  {%
\begin{array}
[c]{ccccccc}%
1.0000 & 0.3333 & 0.3333 & -0.3333 & 0.5774 & 0.5774 & 0.5774\\
0.3333 & 1.0000 & -0.3333 & 0.3333 & 0.5774 & -0.5774 & 0.5774\\
0.3333 & -0.3333 & 1.0000 & 0.3333 & 0.5774 & 0.5774 & -0.5774\\
-0.3333 & 0.3333 & 0.3333 & 1.0000 & 0.5774 & -0.5774 & -0.5774\\
0.5774 & 0.5774 & 0.5774 & 0.5774 & 1.0000 & 0.0000 & 0.0000\\
0.5774 & -0.5774 & 0.5774 & -0.5774 & 0.0000 & 1.0000 & -0.0000\\
0.5774 & 0.5774 & -0.5774 & -0.5774 & 0.0000 & -0.0000 & 1.0000\\
&  &  &  &  &  &
\end{array}
}\right)  .
\]
\end{widetext}
$X$ satisfying the constraint that $X$ is SDSP with non-negative eigenvalues
\cite{SDP2}.

\item {\textbf{For the case $k=3$ to $k=8$} }

After setting up the SDP for finding the Tsirelson bound, we still use the
package named SeDuMi to solve it for both case (a) and (b) with any value of
$k$. The result agrees extremely well with the bound obtained from information
causality up to $\mathcal{O}(10^{-4})$. To be more concrete, the numerical
results are shown below: for case (a) up to $k=8$, we have

\begin{center}%
\begin{tabular}
[c]{|c|c|c|c|c|c|c|}\hline
k & 3 & 4 & 5 & 6 & 7 & 8\\\hline
SDP & 6.9282 & 16.0000 & 35.7771 & 78.3837 & 169.3281 & 362.0387\\\hline
\end{tabular}
\end{center}
This agrees extremely well with the RHS of (\ref{Casea}). Similarly, for case
(b) up to $k=8$, we have

\begin{center}%
\begin{tabular}
[c]{|c|c|c|c|c|c|c|}\hline
k & 3 & 4 & 5 & 6 & 7 & 8\\\hline
SDP & 13.8564 & 32.0000 & 71.5542 & 156.7673 & 338.6562 & 724.0773\\\hline
\end{tabular}

\end{center}
\end{itemize}

It again agrees extremely well with (\ref{Caseb}). Therefore, based on our
numerical simulation, information causality indeed singles out the Tsirelson
bound of a physical theory such as quantum mechanics.


\end{document}